\documentclass[preprintnumbers,showpacs,amsmath,amssymb,prd,floatfix,twocolumn,superscriptaddress,nofootinbib]{revtex4}
\usepackage{graphicx}
\usepackage{epsfig}
\usepackage{bm}
\usepackage{amsfonts}

\begin{document}

\title{Quintessence and phantom dark energy from ghost D-branes}

\author{Emmanuel N. Saridakis }
\email{msaridak@phys.uoa.gr} \affiliation{Department of Physics,
University of Athens, GR-15771 Athens, Greece}

\author{John Ward}
\email{jwa@uvic.ca} \affiliation{Department of Physics and
Astronomy University of Victoria, Victoria, BC, V8P 1A1, Canada}

\begin{abstract}
We present a novel dark energy candidate, based upon the existence
and dynamics of Ghost $D$-branes in a warped compactification of
type IIB string theory. $Gp$-branes cancel the combined BPS
sectors of the $Dp$-branes, while they preserve the same
supersymmetries. We show that this scenario can naturally lead to
either quintessence or phantom-like behaviors, depending on the
form of the involved potentials and brane tension. As a specific
example we investigate the static, dark-energy dominated solution
sub-class.
\end{abstract}

\pacs{95.36.+x, 11.25.Uv, 98.80.-k}
 \maketitle

\section{Introduction}

The theoretical description of the observed universe acceleration
\cite{observations} is one of the challenges of current research.
The simplest way to explain this remarkable behavior (apart form
the sole cosmological constant which leads to the corresponding
problem) is to construct various ``field'' models of dark energy,
using a canonical scalar field (quintessence) \cite{quint}, a
phantom field, that is a scalar field with a negative sign of the
kinetic term \cite{phantBigRip,phant}, or the combination of
quintessence and phantom in a unified model named quintom
\cite{quintom}. However, the arbitrary consideration of additional
scalar fields (which may even have non-conventional kinetic terms
inserted by hand) should be constrained by the fact these extra
scalars to be neutral under all the standard model symmetries, and
thus not introducing additional fifth forces. This non-trivial
requirement led many authors to the alternative direction of
modifying gravity itself \cite{ordishov}, with a promising attempt
in these lines being perhaps the recent developments in
Ho\v{r}ava-Lifshitz gravity \cite{Horava}, a power-counting
renormalizable, Ultra-Violet (UV) complete gravitational theory
(although there may well be problems with the theory due to
additional degrees of freedom becoming strongly coupled in the
Infra-Red).

On the other hand, constructions arising from string theory are
hard to result in dark energy phenomenology consistent with
observations, purely using the closed string sector. For more
details see the review \cite{Grana:2005jc}. However, cosmological
dynamics driven by the open string sector through dynamical
$Dp$-branes, which is the basic idea of the so-called Dirac Born
Infeld (DBI) formalism, has led to interesting successes, mainly
in inflationary paradigms \cite{Alishahiha:2004eh,
Kecskemeti:2006cg,Lidsey:2007gq}. After inflation the universe
lives on branes that wrap various cycles within the compact space,
and in this sense the GUT or Electro-Weak phase transition can be
manifested through a geometric fashion. Thus, dark energy does
present a dynamical nature, retaining additionally a form of
geometric origin. Quantitatively, the tight constraints from WMAP
$5$ year data set \cite{Komatsu:2008hk} on the model parameters
have led DBI-models to more complex versions, including multiple
fields \cite{Easson:2007dh}, multiple branes \cite{Cai:2008if,
Cai:2009hw, Thomas:2007sj}, wrapped branes \cite{Becker:2007ui} or
monodromies \cite{Silverstein:2008sg}. Finally, the phase-space
analysis of a solitary $D3$-brane moving through a particular
warped compactification of type IIB was done \cite{singlebrane},
while the generalization to multiple and partially wrapped branes
has been performed in \cite{Gumjudpai:2009uy}.

In the present work we are interested in constructing a
DBI-scenario based on Ghost $D$-branes, that is $Dp$-branes that
have a $\mathbb{Z}_2$ symmetry acting to flip the signs of the
$NS$-$NS$ and $RR$ sectors. Such a consideration is more robust
than the naive and ambiguous use of the prototypical non-BPS
$D3$-brane action, albeit with the wrong sign kinetic term. In
addition, although in a typical flux compactification of type II
string theory down to four-dimensions one must introduce negative
tension objects called Orientifolds (in order to cancel the
$D3$-brane charge associated with the closed string fluxes, and to
project out various string states breaking half of the bulk
supersymmetry so that the vacuum retains an $\mathcal{N}=1$
structure), the existence of Ghost branes could possibly negate
the need for Orientifolds. As we show, in such a ghost $D$-brane
scenario we can naturally acquire an effective dark energy
behaving either as quintessence or as phantom.

The plan of the work is as follows: In section \ref{model} we
present the formalism of $Dp$-branes, used in dark energy
scenarios. In section \ref{Ghostd} we extend it, introducing the
concept of ghost D-branes, and we extract the dark-energy
equation-of-state parameter. In section \ref{cosmimpl} we
investigate its general features, examining the conditions for the
appearance of quintessence or phantom behavior, while in section
\ref{staticde} we perform an explicit phase-space analysis of the
static, dark-energy dominated, solution sub-class. Finally, our
results are summarized in section \ref{conclusions}.

\section{$Dp$-brane action}
\label{model}

The open string sector of type II string theory is usually
governed by the DBI action, governing the low energy fluctuations
of such strings attached to a $Dp$-brane. For $N$ coincident
branes, the world-volume symmetry is enhanced from $U(1)$ to
$U(N)$, and the scalar fluctuations are then promoted to matrices
obeying a Lie structure. This is similar to the induced
non-commutative world-volume theory on a single $Dp$-brane when we
turn on a non-trivial $B$-field.

String scattering calculations indicate that there is a particular
trace prescription required in order to account for the full
string cross-section, which is given by the symmetrized average
over all possible orderings of the Lie-algebra valued objects
\cite{Myers:1999ps}. Much like the single $Dp$-brane action, one
can sum the relevant terms into non-linear form (although only
valid up to $\mathcal{O}(\alpha')^3$) to partially reconstruct the
non-Abelian theory using the effective action \cite{Myers:2003bw, Brecher:2004qi}:
\begin{widetext}
\begin{equation}
S =- T_p \int d^{p+1}\xi \rm{STr} \left(e^{-\phi}
\sqrt{-\rm{det}(\mathcal{P}[E_{ab}+
E_{ai}(Q^{-1}-\delta)^{ij}E_{jb}+ \lambda
F_{ab})])}\sqrt{\rm{det}Q^i_j}\right) \pm \mu_p \int \rm{STr}
\mathcal{P}[e^{i\lambda i_{\phi}i_{\phi}}\sum C^{(n)}
e^{B}]e^{\lambda F},
\end{equation}
\end{widetext}
 where
\begin{eqnarray}
&&\lambda = 2\pi \alpha'\nonumber\\
 &&E_{ab} = G_{ab}+B_{ab}\nonumber\\
 && Q^i_j = \delta^i_j + i\lambda [\psi^i, \psi^k]
E_{kj}.
\end{eqnarray}
In the expressions above, $\lambda$ is the inverse of the
$F$-string tension, and $\alpha'$ is the square of the
string-length - the fundamental length scale in our theory. The
scalar $\psi^i$ is related the space-time embedding $X^i=\lambda
\psi^i$ and finally $B^{(2)}$ is the $NS$-$NS$ gauge potential,
which we will ignore in the following. Moreover $\mathcal{P}$
denotes the pullback operator acting on the bulk space-time tensor
fields, and $\psi^i$ are the scalar field fluctuations where
$i=(p+1)\ldots 9$. In the $RR$-sector we see the introduction of
the interior derivative $i_{\psi}$, whose action on an $n$-form is
\begin{equation}
i_{\psi}i_{\psi} C^{(n)} = \frac{1}{2} [\psi^i, \psi^j]C_{ji}^{(n)}.
\end{equation}
The presence of the interior rather than exterior derivative
allows the $Dp$-brane to couple to gauge potentials of higher
order, such as the $(p+3), (p+5)$ forms. This suggests that there
is a transmutation (or dielectric) effect where the $Dp$-brane can
blow up into a $D(p+2)$-brane through higher order terms in the
expansion of the Chern-Simons action \cite{Constable:1999ac,
Myers:2003bw}. A concrete example of this effect is when $N$
$D3$-branes blow up into a solitary $D5$-brane via the formation
of a fuzzy $S^2$, more commonly referred to as the Myers effect.
If the scalar fields transform under an appropriate representation
of a higher dimensional gauge group, then the $D3$-branes can be
polarized into higher dimensional branes in an analogous fashion
through the extended Myers effect \cite{Myers:2003bw,
Hyakutake:2004rh}\footnote{Although there exists a different
action, proposed by Tseytlin \cite{Tseytlin:1997csa}, which does
not admit such an effect.}. For example, if the scalars transform
under irreducible representations of the $n$-fold tensor product
of $SO(5)$, then the branes orient themselves along a fuzzy $S^4$
to form a configuration of $n$ $D7$-branes
\cite{Constable:2001ag}. The construction of odd-dimensional fuzzy
sphere solutions is actually non-trivial and requires the
introduction of spinorial representations
\cite{Papageorgakis:2006ed} of $SO(2k)$, where $k \in \mathbb{Z}$.

One important simplification to the above action is when we
consider the large-$N$ limit, as the STr operation reduces to a
trace (up to $1/N$ corrections). The reason why this limit is
important can be understood when one considers the dual
description of the brane configuration. Recall that the Myers
effect describes lower-dimensional branes being polarized into
higher-dimensional configurations via a fuzzy sphere. This means
that there is a dual description of the Myers effect in terms of a
higher-dimensional (spherical) brane with world-volume flux. More
concretely we see that that (assuming the scalars lying in
irreducible representations of $SO(3)$) $N D3$-branes are dual to
a single $D5$-brane wrapped on $S^2 \times R^3$ with $N$ units of
flux through the $S^2$ \cite{Thomas:2006ac}. Duality in this sense actually
means that the effective actions are identical, provided that the
$U(1)$ flux on the $D5$-brane is large. The above statements are
all assumed to be true in a curved background, although the
required string scattering calculations are difficult to be
computed to the necessary order and therefore a direct check is
not possible. However given the prevalence of such dualities in
string theory, one can be reasonably confident that the statement
is correct.

The most general cosmological backgrounds in type II string theory
can be written in the following form \cite{McAllister:2007bg}
\begin{equation}\label{eq:metric}
ds^2 = h^2 (\rho) ds_4^2 + h^{-2}(\rho)(d\rho^2 + \rho^2
ds_{X_5}^2),
\end{equation}
where $h$ is the warp factor, which is a function of $\rho$ - a
warped throat that is fibred over some five-dimensional manifold
$X_5$, and the four-dimensional metric takes the usual FRW-form.
For concreteness we will specialize to the case of type IIB string
theory, where the throat can be generated by threading $D3$-brane
flux through a compact three-cycle. Moreover, since the dilaton is
constant in these backgrounds, the Einstein frame and String
frames coincide. We will also assume that our theory consists of
$N D3$-branes which are oriented parallel to the $(3+1)$-large
dimensions and that the scalars are homogeneous, transforming
under irreducible representations of $SO(3) \sim SU(2)$. The
resulting action for $N$ coincident $D3$-branes can be written as
\cite{Thomas:2007sj}\footnote{For $\bar{D}3$-branes we would require the
second term to take the positive sign.} {\small{
\begin{eqnarray}
\label{action0}
&&\!\!\!\!\!\!\!\!\!\!S= -T \int d^4 \xi N
\sqrt{-g_4} \left[h^{4}\sqrt{1-h^{-4}\lambda^2
\hat{C}\dot{R}^2}\sqrt{1+4\lambda^2 \hat{C}h^{-4}R^4}-\right.\nonumber\\
&&\ \ \ \ \ \ \ \ \ \ \ \ \ \ \ \ \ \ \ \ \ \ \ \ \ \ \ \ \ \ \ \
\ \ \ \ \ \ \ \ \ \ \ \ \ \ \ \ \ \ \ \ \ \ \ \left.- h^4+V(R)
\right],
\end{eqnarray}}}
with $T$ the warped, positive-definite brane tension. The radius
of the fuzzy sphere is defined in terms of the geometric radius
$\rho$ via
\begin{equation}
R^2 = \frac{\rho^2}{\lambda^2 \hat{C}},
\end{equation}
and $\hat{C}$ is the quadratic Casimir of $SU(2)$, namely $\hat{C}
= N^2-1$. We have also included a scalar potential contribution
arising from the interaction of the $D3$-branes with the closed
string background. We remind the reader that at large $N$ this
action is precisely the same as that arising from a single wrapped
$D5$-brane with $N$ units of $U(1)$ flux.

It is convenient to use the field redefinition $\phi =
\rho/\sqrt{T}$, with $\rho$ the induced world-volume scalar coming
from the background in the string frame \cite{Thomas:2007sj}.
However, it has (mass) dimension $-1$ and thus since we desire to
write all the fields with canonical mass terms we have to redefine
the involved world-volume scalars. Following these lines, the
action (\ref{action0}) can be re-written in the generalized form
\begin{equation}
S = -\int d^4 \xi \sqrt{-g_4} \left[T(\phi)W(\phi)
\gamma^{-1}-T(\phi) + V(\phi) \right],
\end{equation}
where $\gamma = [1-\dot{\phi}^2/T(\phi)]^{-1/2}$ is the usual
generalization of the relativistic factor. The cosmological
consequences of such an action have been discussed elsewhere
\cite{Thomas:2007sj} and we refer the interested reader there for
more details.

The appearance of the positive-definite function $W(\phi)$, which
generalizes the aforementioned action comparing to the usual
$W(\phi)\equiv1$ case, can be theoretically justified
\cite{Gumjudpai:2009uy}, since if $N$ multiple coincident branes
are present then the world-volume field theory is a $U(N)$
non-Abelian gauge theory and this ``potential'' term is simply a
reflection of the additional degrees of freedom
\cite{Myers:1999ps}. Additionally, this configuration is related
to a $D5$-brane, wrapping a two-cycle within the compact space and
carrying a non-zero magnetic flux along this cycle. On the other
hand, the positive-definite effective potential $V(\phi)$ accounts
for the possible open or closed string interactions. Its precise
form depends upon the number of additional branes and geometric
moduli, the number of non-trivial cycles in the compact space, the
choice of embedding for branes on these cycles, the coupling of
the brane to any background RR-form fields, the contribution from
higher dimensional bulk forms \cite{Langlois:2009ej} etc. Finally,
note that using the above generalized form for the action, allows
us to interpolate between a single $D3$-brane (taking
$W(\phi)\to1$) and the multi-brane, or wrapped $D5$-brane (where
$W(\phi) > 1$) solutions.

\section{Ghost $D$-brane cosmology}
\label{Ghostd}

It is well  established that $Dp$-branes are not the only
hypersurfaces within string theory. There are also Orientifold
$Op$-planes which have negative tension and reduced charge
(compared to the $Dp$-branes) \cite{Dabholkar:1997zd}. Their role is vital in
flux compactifications of type II string theory, since they cancel
global flux tadpoles and also break one half of the residual
supersymmetries. There exists another type of extended object,
which has been dubbed a Ghost-brane \cite{Okuda:2006fb}, that we
will briefly describe using the boundary state formalism, which is
the most appropriate for the CFT description of $Dp$-branes.

 The
bosonic sector of a BPS $Dp$-brane is represented by a boundary
state of the form
\begin{equation}
|D> = |D>_{NS NS}+|D>_{RR},
\end{equation}
where $|D>$ represents the full $Dp$-brane state.  It was shown in
\cite{Okuda:2006fb, Terashima:2006qm} that one can define an
analogous (BPS) Ghost-brane state (which we will denote by a
$Gp$-brane) through the introduction of an operator $g$:
\begin{equation}
|G> = |gD> = -|D>,
\end{equation}
such that the ghost state precisely  cancels the combined BPS
sectors of the $Dp$-brane. Since the $Gp$-brane state preserves
the overall relative sign of the two different sectors, it must
also preserve the same supersymmetries as the $Dp$-brane. This
makes it a distinct object, and it should not be confused with the
$\bar{D}p$-brane - which has a boundary state of the form
\begin{equation}
|\bar{D}> = |D>_{NS NS} - |D>_{RR}.
\end{equation}
The formalism implies that the $Gp$-sector exactly cancels the
$Dp$-sector. This means that a theory consisting of $N$ coincident
$Dp$-branes and $M$ coincident $Gp$-branes can be described in two
equivalent ways; either by $(N-M)$ $Dp$-branes or by $(M-N)$
$Gp$-branes. The corresponding world-volume theory is given by a
$U(N) \times U(M)$ gauge theory, which is enhanced to $U(N|M)$ as
the two groups of branes are brought together.  Importantly, this
means that when $N=M$ the resulting solution is simply space-time
with no $Dp$-branes. In this way we see that the ghost brane can
screen the $Dp$-brane, and a useful consequence of this screening
was employed in AdS/CFT framework in \cite{Evans:2006eq}.

Since the $Gp$-brane is simply minus the standard  $Dp$-brane
state, one sees that the effective world-volume theory for the
$Gp$-brane is also of DBI form, albeit with an additional sign
change in the definition of the tension. Thus, for multiple
coincident $G3$-branes, we expect the effective theory to be well
described by the action
\begin{equation}
\label{action}
 S = \int d^4 \xi \sqrt{-g_4} \left[T(\phi)W(\phi)
\gamma^{-1}-T(\phi) - V(\phi) \right],
\end{equation}
where we have embedded the branes in the warped background
(\ref{eq:metric}). Note that in the non-relativistic expansion of
this action, the kinetic term will have the wrong sign, implying
phantom-like behavior for the scalar fluctuations. This suggests
that the world-volume theory tends to anti-gravitate, rather than
gravitate.

We stress that the $G3$-brane theory is different from previously
proposed phantom models based on the DBI-action \cite{Hao:2003ib},
which have been constructed in the light of the non-BPS action
proposed by Sen as an effective description of tachyon
condensation \cite{Sen:1999md}. The models in this class have the
wrong sign kinetic term \emph{inside} the square-root structure,
in contrast to that of the ghost action. Furthermore, that
sign-change can only be inserted by hand \cite{Vikman:2004dc},
since the world-volume metric is induced from the background
geometry, and it is unlikely to contain a sub-manifold where the
sign changes in a continuous fashion \footnote{Leaving aside any
issues concerning type II* string theory \cite{typeII*} for the
moment.}. Additionally, the non-BPS action can only couple to any
of the bulk $RR$-form fields through terms of the form $dT \wedge
C^{(3)}$, which are typically zero according to our assumptions.
Therefore, it seems unlikely that such boundary states can be
stable within the full theory. On the other hand, the ghost branes
are supersymmetric and do couple to the bulk form-fields,
suggesting that they constitute actually stable states within the
theory.

Let us now focus on the cosmological consequences of the scenario
at hand. Assuming that the scalar is time-dependent, one reads off
the diagonal components of the energy momentum-tensor in the usual
fashion:
\begin{eqnarray}
\label{rhophi}
&&\rho_{\phi} = T(\phi)[1-W(\phi)\gamma] + V(\phi)\\
\label{pphi} && p_{\phi} = T(\phi)[W(\phi)\gamma^{-1}-1]-V(\phi).
\end{eqnarray}
Thus, since in DBI-constructions $\phi$ is responsible for dark
energy, we can define its equation-of-state parameter as:
\begin{equation}
\label{wphi}
w_{\phi} = \frac{p_{\phi}}{\rho_{\phi}} =
\frac{T(\phi)[W(\phi)\gamma^{-1}-1]-V(\phi)}{
T(\phi)[1-W(\phi)\gamma] + V(\phi)}.
\end{equation}
As can be deduced from  expression (\ref{wphi}), the dark-energy
equation-of-state parameter can present quintessence or phantom
behavior, depending on the choice of scalar potential and
background.

The Friedmann equations arising from action (\ref{action}) write:
\begin{eqnarray}
\label{Fr1}
H^2&=& \frac{1}{3 M_p^2}(\rho_M + \rho_{\phi}),\\
\label{Fr2}
 \dot{H}&=& -\frac{1}{2 M_p^2} \left[\rho_M+p_M +
\rho_\phi+p_\phi \right]\nonumber\\
&=& -\frac{1}{2 M_p^2} \left[\rho_M+p_M -\gamma W(\phi)
\dot{\phi}^2 \right],
\end{eqnarray}
with $H\equiv \dot{a}/a$ the Hubble parameter. Additionally,
variation of the action (\ref{action}) with respect to $\phi$
leads to the equation of motion for the scalar field, namely:
\begin{eqnarray}
3 H W(\phi)\, \gamma \dot{\phi}+W(\phi) \gamma^3
\ddot{\phi}-V_\phi(\phi)+W_\phi(\phi)T(\phi)\gamma +\nonumber\\
+\frac{T_\phi(\phi)}{2}\left[ W(\phi) \gamma(3-\gamma^2)-2 \right]
=0,\
\end{eqnarray}
where the subscript $_\phi$ denotes differentiation with respect
to $\phi$. This equation is the generalization of the Klein-Gordon
one in the DBI framework, and using (\ref{rhophi}),(\ref{pphi}) it
can be written in the usual form
$\dot{\rho}_\phi+3H(\rho_\phi+p_\phi)=0$. Finally, the
corresponding equation of motion for  matter  writes
$\dot{\rho}_M+3H(\rho_M+p_M)=0$.

\section{Cosmological implications: the general case}
\label{cosmimpl}

In the previous section we introduced the concept of ghost
D-branes, and we extracted the dark-energy equation-of-state
parameter of this ghost version of DBI scenario. Thus, we can now
investigate the various cosmological possibilities, trying to
remain sufficiently general. We mention that we desire to explore
the general features of $w_\phi$ for possible forms of the
involved tension and potentials, without examining in detail the
equations of motion. As we see, although a full dynamical
investigation would be interesting, this basic ``kinematical''
study is sufficient to qualitatively reveal the novel features of
the ghost $D$-brane model.

Let us first consider the scenario where no scalar potential is
present, that is study solely the brane action. In this case the
equation of state reduces to
\begin{equation}
w_{\phi} = \frac{1}{\gamma} \left[
\frac{W(\phi)-\gamma}{1-W(\phi)\gamma}\right].
\end{equation}
For a general $W(\phi)$, in the regime where $\gamma >>1$ we see
that the equation of state is typically zero, unless there are
divergences in $W$, which is not the case if we desire our model
to be physical. On the other hand, in regions where $W(\phi)$ is
dominant we find that $w_{\phi} \sim -1/\gamma^2$ and therefore it
is negative-definite (although small). Similarly, if $W(\phi) = 1$
then $w_{\phi} = 1/\gamma$ which is positive-definite although
typically small. Note that physical solutions imply $W(\phi) \ge
1$, however if we treat the action phenomenologically and assume
smaller values for $W(\phi)$ then we find solutions where
$w_{\phi} \to 0$ from above after being initially large. In
conclusion, we observe that the possible solution space is quite
large even without a scalar potential.

This preliminary phenomenology suggests that $w_{\phi}$ could
cross the $-1$-bound. In particular, the equation of state would
become phantom if
\begin{equation}
\frac{W(\phi) [1-\gamma^2]}{[1-W(\phi)\gamma]} < 0.
\end{equation}
However, this condition cannot be met physically, and thus we
conclude that the brane action alone cannot generate phantom
dynamics.

Let us now turn on the scalar potential term $V(\phi)$. A first
simple solution subclass would be to consider $T(\phi)=0$, where
we obtain  $w_{\phi}=-1$  recovering the case of pure de-Sitter
expansion. In the general case of non-zero potential and tension
terms, but with $V(\phi)\gg T(\phi)$, we can expand (\ref{wphi})
in Taylor series acquiring:
\begin{equation}
w_{\phi} \approx -1 + \frac{T(\phi)  }{V(\phi)}
\frac{W(\phi)(1-\gamma^2)}{\gamma}+\ldots,
\end{equation}
neglecting higher order  terms. Therefore, in the relativistic
regime ($\gamma^2>>1$) the correction term will be
negative-definite, leading to the realization of the phantom
phase. We mention that this phantom realization is obtained
naturally from a large solution subclass of the model.
Additionally, it is not the only combination of possibilities
which lead to phantom behavior, but just a simple example. These
features reveal that the use of ghost D-branes does lead to
quintessence and phantom realization, depending on the specific
forms of the potential-like terms and of the tension in the
effective action.

Another class of solutions will occur when we have $T(\phi) \gg
V(\phi)$,  which upon performing the Taylor expansion leads to
\begin{equation}
\label{wquint}
 w_{\phi} \approx
\frac{\gamma-W(\phi)}{\gamma[\gamma W(\phi)
-1]}\left\{1+\frac{V(\phi)}{T(\phi)}\frac{W(\phi)(\gamma^2-1)}{[\gamma
W(\phi)-1][\gamma-W(\phi)]}\right\}
\end{equation}
at leading order, and therefore it is highly dependent on the
particular background field-parametrization. For initially static
configurations ($\dot{\phi}=0$, i.e $\gamma=1$) we recover the
usual result $w_{\phi} \approx -1$, and therefore the static brane
mimics the cosmological constant. As the velocity of the brane
increases we again find that $w_{\phi} \to 0$ along the asymptotic
branch. On the other hand, if $W(\phi)\gg1$ then the equation of
state tends to $-1/\gamma^2$ and therefore will relax to zero from
below. Finally, imposing $W(\phi)=1$, that is considering the
single brane case, the resulting equation-of-state parameter tends
to zero from above as the velocity term increases, as can be seen
from (\ref{wquint}). Note that since $\gamma\ge1$, all the cases
of the regime $T(\phi) \gg V(\phi)$ present a quintessence
behavior with $w_\phi\ge-1$.

As we have mentioned, in the present work we are interested in
exploring the general qualitative features of the
equation-of-state parameter of ghost $D$-brane scenario. We have
not extracted the equations of motion, studying just $w_\phi$ as a
function of $T(\phi)$, $W(\phi)$, $V(\phi)$ and $\gamma$, which is
itself a function of $\phi$ and $\dot{\phi}$. Therefore, for given
$T(\phi)$, $W(\phi)$, $V(\phi)$, the dependence of $w_\phi$ on
$\gamma$ provides qualitative information about the phase-space
structure. A first observation is that (\ref{wphi}) possesses a
singularity at
\begin{equation}
\label{gammac}
 \gamma_c=\frac{1}{W(\phi)}\left[1+\frac{V(\phi)}{T(\phi)}\right].
\end{equation}
According to the specific choice of $T(\phi)$, $W(\phi)$,
$V(\phi)$ and of initial conditions, a particular universe
evolution (i.e a particular orbit of $\gamma(\phi,\dot{\phi})$ in
the $(\phi,\dot{\phi})$-plane) can remain either to one or to the
other regime, tend asymptotically into the singularity, or even
cross it. Such singularities are common in field dark energy
models, especially in phantom ones, and they correspond to Big Rip
\cite{phantBigRip,BigRip} or to realization of a cosmological
bounce \cite{Bounce}. Finally, note that if $\gamma_c$ turns out
to be less than $1$, that is unphysical, then the specific model
is free of such behaviors, independently of the initial
conditions.

\begin{figure}[ht]
\begin{center}
\mbox{\epsfig{figure=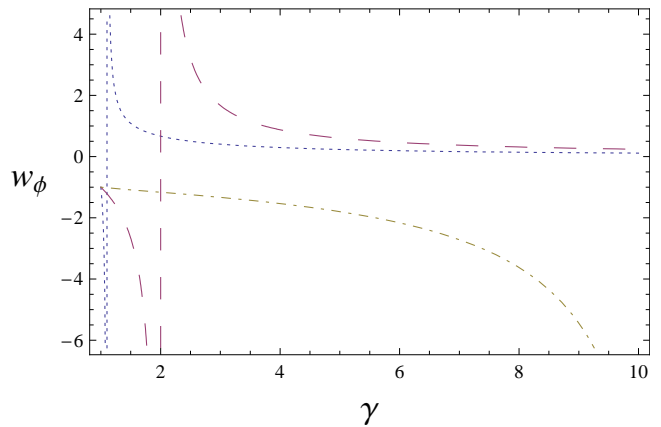,width=8.5cm,angle=0}} \caption{{\it
The dark energy equation-of-state parameter $w_{\phi}$ as a
function of the generalized relativistic factor $\gamma$, for
fixed $W(\phi)=1$. The curves correspond to $V(\phi)/T(\phi)=0.1$
(dotted), $1$ (dashed) and $10$ (dotted-dashed) respectively.
 }} \label{fig1}
\end{center}
\end{figure}

In order to provide a more transparent picture of the obtained
cosmological behavior, in fig.~\ref{fig1}  we present the solution
space for the simple scenario of fixed $V(\phi)/T(\phi)$, imposing
$W(\phi)=1$ (corresponding to the single brane model).
As we observe, for $V(\phi)/T(\phi) \ll 1$ there is a singularity
at $\gamma_c=1.1$, thus in almost the whole phase-space the
equation-of-state parameter is quintessence-like and in particular
it is positive-definite and tends to zero asymptotically. For
larger values of $V(\phi)/T(\phi)$ (i.e when the scalar potential
effect is enhanced comparing to that of the tension) we observe
the singularity at a specific $\gamma_c$, which in this special
subclass (fixed $V(\phi)/T(\phi)$ and $\phi$-independent $W$) is
constant, i.e $\phi$-independent.  For values of $\gamma <
\gamma_c$ phantom behavior is realized, while for $\gamma
> \gamma_c$ one finds quintessence-like evolution with
$w_{\phi} \to 0$ from above.

\begin{figure}[ht]
\begin{center}
\mbox{\epsfig{figure=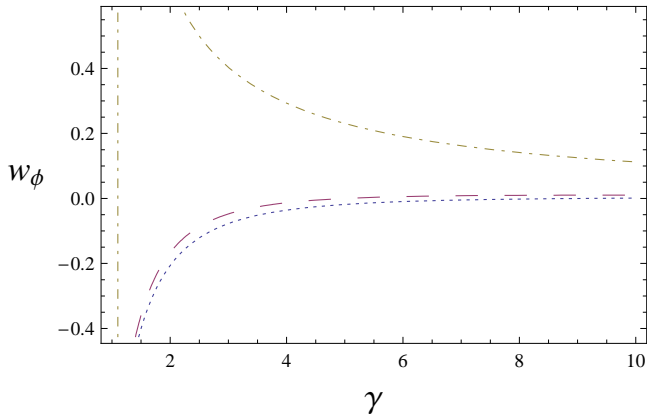,width=8.5cm,angle=0}} \caption{{\it
The dark energy equation-of-state parameter $w_{\phi}$ as a
function of the generalized relativistic factor $\gamma$, for
fixed $W(\phi)=10$. The curves correspond to $V(\phi)/T(\phi)=0.1$
(dotted), $1$ (dashed) and $10$ (dotted-dashed) respectively.
 }} \label{fig2}
\end{center}
\end{figure}

Let us now consider the same subclass of fixed $V(\phi)/T(\phi)$,
but setting $W(\phi)=10$. This scenario can be obtained in a class
of string theory backgrounds with $G3$-branes or a $G5$-brane with
flux. In fig.~\ref{fig2} we depict the corresponding
$w_\phi$-behavior as a function of $\gamma$.
In this case, for small values of $V(\phi)/T(\phi)$ the value of
$\gamma_c$ is unphysical. Therefore, the resulting trajectories
are quintessence-like, and the model is singularity-free
independently of the initial conditions. For larger values of
$V(\phi)/T(\phi)$, $\gamma_c$ becomes physical, with $\gamma <
\gamma_c$ leading to phantom behavior and $\gamma> \gamma_c$ to
quintessence-like one with $w_{\phi} \to 0_{+}$ or $w_{\phi} \to
0_{-}$.

In summary, from this simple solution-subclass we observe an
interesting $w_\phi$-behavior.  We mention that in principle,
$W(\phi)$ and $T(\phi)$ are determined by the supergravity
background and can have various forms, while $V(\phi)$ can be more
arbitrary since it arises from the interactions of the open/closed
string sector which are difficult to compute in general. Clearly,
considering more general scenarios, with various $T(\phi)$ and
$V(\phi)$ or/and not constant $W(\phi)$, the resulting
cosmological behavior can be significantly richer.

\section{Static dark-energy-dominated solutions}
\label{staticde}

Having discussed qualitatively the cosmological behavior of the
model at hand, it would be interesting to perform a systematic
investigation of the various cosmological solution sub-classes. In
particular, we desire to study the cosmologically important
scenario of static solutions characterized by complete dark energy
domination. We examine whether there exist late-time attractor
solutions, and if they do exist to determine their observable
features, that is the dark-energy equation-of-state parameter and
density parameter. Furthermore, we want to extract information
about the intermediate-time behavior, that is the evolution
towards the aforementioned late-time attractors, since such an
investigation could also leave imprints in observables related to
the cosmological past.

 As usual, we will first transform the cosmological system
into its autonomous form \cite{expon}: $
\dot{\textbf{X}}=\textbf{f(X)} $, where $\textbf{X}$ is the column
vector constituted by the (suitably defined) dimensionless
variables and \textbf{f(X)} the corresponding column vector of the
autonomous equations, and we extract its critical points
$\bf{X_c}$ satisfying $\dot{\bf{X}}=0$. Then, in order to
determine the stability properties of these critical points, we
expand   around $\bf{X_c}$, setting $\bf{X}=\bf{X_c}+\bf{U}$ with
$\textbf{U}$ the perturbations of the variables considered as a
column vector. Thus, for each critical point we expand the
equations for the perturbations up to the first order as $
\dot{\textbf{U}}={\bf{\Xi}}\cdot \textbf{U}, $ where the matrix
${\bf {\Xi}}$ contains the coefficients of the perturbation
equations. Thus, for each critical point, the eigenvalues of ${\bf
{\Xi}}$ determine its type and stability.

Defining the dimensionless variables
\begin{equation}
X = \frac{\phi}{M_p}, \hspace{0.5cm} Y =
\frac{\dot{\phi}}{\sqrt{T}},
\end{equation}
 the equations of motion reduce to the following set of
equations
\begin{eqnarray}
\dot{X} &=& \frac{Y\sqrt{T}}{M_p} \\
\dot{Y} &=& \frac{V_\phi}{W\gamma^3 \sqrt{T}}-
\frac{T_{\phi}}{\sqrt{T}}\left[\frac{(3-\gamma^2)}{2\gamma^2}+\frac{Y^2}{2}-\frac{1}{W\gamma^3} \right] \nonumber \\
&-&
\frac{W_{\phi}}{W}\frac{\sqrt{T}}{\gamma^2}-\frac{\sqrt{3}Y}{\gamma^2
M_p}\sqrt{T(1-W\gamma)+V}
\end{eqnarray}
where we have set $\rho_M=0=p_M$ since we are investigating the
complete dark-energy dominated scenario. Furthermore, in terms of
the dimensionless variables, the dark energy equation-of-state
parameter (\ref{wphi}) writes:
\begin{eqnarray}
\label{wphiXY}
w_\phi=\sqrt{1-Y^2}\left[\frac{W(X)\sqrt{1-Y^2}-1}{\sqrt{1-Y^2}-W(X)}\right].
\end{eqnarray}

Since we are interested in static late-time attractors, that is
possessing $\dot{\phi}=0$, the corresponding critical points are
of the form $(X_c,0)$. Thus, linearized perturbations ($X=X_c+
\delta X$, $Y=0+\delta Y$) lead to
\begin{widetext}
\begin{eqnarray}
\dot{\delta X} &=& \frac{\sqrt{T}\delta Y}{M_p} \nonumber \\
\dot{\delta Y} &=& \frac{\delta X}{ W \sqrt{T}}
\left\{V_\phi\left\lbrack\frac{V_{\phi\phi}}{V_\phi}-\frac{W_\phi}{W}-
\frac{T_\phi}{2T}\right\rbrack-W_\phi
T\left\lbrack\frac{W_{\phi\phi}}{W_\phi}-\frac{W_\phi}{W}\right\rbrack-\frac{T_\phi}{2}\left\lbrack\frac{T_{\phi\phi}}{T_\phi}+
\frac{W_\phi}{W}-\frac{T_\phi}{2T}\right\rbrack \right\} \nonumber \\
&+&\delta Y \left\{-3 H_0 -\frac{T_\phi}{
\sqrt{T}}+\frac{3V_\phi}{2W
\sqrt{T}}-\frac{W_\phi\sqrt{T}}{W}\left\lbrack1-\frac{T_\phi}{2T}\right\rbrack-
\frac{T_\phi(W-2)}{4W\sqrt{T}} \right\}\nonumber \\
&=& \alpha \delta X + \beta \delta Y,\label{eqsfull}
\end{eqnarray}
\end{widetext}
where all the derivative terms on the right hand side are
evaluated at $X=X_c$, and $H_0$ stands for the value of the Hubble
parameter (given by (\ref{Fr1}) and (\ref{rhophi})) calculated at
$X=X_c$. Thus, the corresponding stability matrix reads
\[{\bf{\Xi}}= \left[ \begin{array}{cc}
0 & \frac{\sqrt{T}}{M_p}  \\
\alpha & \beta
\end{array} \right],\]
and its eigenvalues are $\lambda_{\pm} = \frac{1}{2} \left(\beta
\pm \sqrt{\beta^2+4\frac{\alpha \sqrt{T}}{M_p}} \right)$.
 Requiring negativity of the eigenvalue real part (which
corresponds to stability of the corresponding fixed point) we
result to the constraint
\begin{eqnarray}
V_\phi\left\lbrack\frac{V_{\phi\phi}}{V_\phi}-\frac{W_\phi}{W}-
\frac{T_\phi}{2T}\right\rbrack-W_\phi
T\left\lbrack\frac{W_{\phi\phi}}{W_\phi}-\frac{W_\phi}{W}\right\rbrack-\nonumber\\
-\frac{T_\phi}{2}\left\lbrack\frac{T_{\phi\phi}}{T_\phi}+
\frac{W_\phi}{W}-\frac{T_\phi}{2T}\right\rbrack < 0.
\label{staticstability}
\end{eqnarray}
In the following we explore the general features of this stability
condition, for various cases of the involved potentials and
tension.

We first consider a solution where $W(\phi)=T(\phi)=$ const., and
therefore condition (\ref{staticstability}) reduces to
\begin{equation}
V_{\phi\phi}<0,
\end{equation}
evaluated at the critical value of $X=X_c$. This expression
(together with the potential positivity) imposes tight
restrictions on the form of the potential, if we desire to obtain
a late-time attractor. In particular, it requires potentials where
the field is initially localized near $\phi \sim 0$ and rolls to
larger values (analogous to the small-field models of inflation).
Candidates are therefore $V \sim V_0 / \cosh(\xi \phi)$, $V=V_0
\cos(\upsilon \phi)$ and $V\sim V_0 - m^2 \phi^2$.

\begin{figure}[ht]
\begin{center}
\mbox{\epsfig{figure=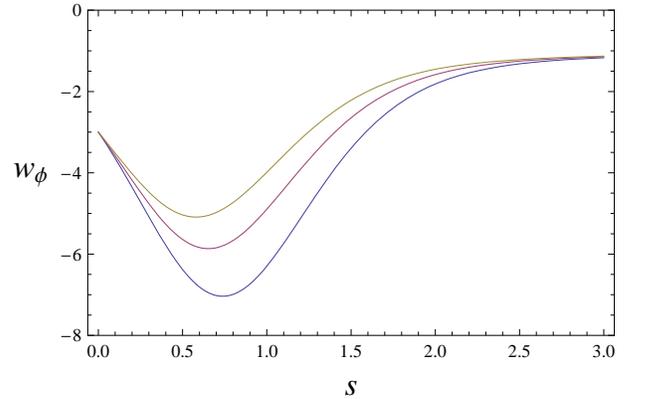,width=8cm,angle=0}} \caption{{\it
$w$-evolution for the potential $V\sim V_0 \phi^2/\phi_0^2$, in
terms of the variable $s=t/\phi_0$, for $M_p=V_0=1$ and
$W(\phi)=T(\phi)=1$. The top curve corresponds to $\phi_0=300$,
the middle curve to $\phi_0 = 250$, and the bottom curve to
$\phi_0=200$. }} \label{tight}
\end{center}
\end{figure}

In order to provide an explicit example of this sub-class, we
consider the potential $V\sim V_0 \phi^2/\phi_0^2$. Transforming
to the variables $\phi=\phi_0\,\delta X, t=\phi_0s$, where
$\phi_0$ is a reference field position,  we can write the
equations of motion (\ref{eqsfull}) in dimensionless form
\begin{eqnarray}
\frac{d(\delta X)}{ds} &=& \delta Y\sqrt{T} \\
\frac{d(\delta Y)}{ds} &=& \frac{2\, \delta X}{\gamma^3
\sqrt{T}}-\frac{\sqrt{3}\,\delta
Y}{\gamma^2}\sqrt{T(1-\gamma)+(\delta X)^2} \nonumber.
\end{eqnarray}
The usual approach is to depict the phase-space plots in the
$(X,Y)$ plane, and show the convergence to a stable fixed point.
However, it is more transparent to depict the evolution of $w$ in
terms of the variable $s$. Thus, convergence of the system to a
static late-time attractor $(X_c,0)$ means convergence of $w$ to
$-1$ (as can be immediately seen from (\ref{wphiXY}) setting
$Y\equiv Y_c=0$). In  fig.~\ref{tight} we depict $w$-evolution,
for $M_p=V_0=1$, $W(\phi)=T(\phi)=1$, and for various
$\phi_0$-values.
 As we can see, the system presents phantom
behavior, going asymptotically to the cosmological-constant
universe. Additionally, we see that the location of the (global)
minimum is shifted to earlier values of $s$ as we increase
$\phi_0$, and it also is closer to the cosmological-constant
equation of state.

\begin{figure}[ht]
\begin{center}
\mbox{\epsfig{figure=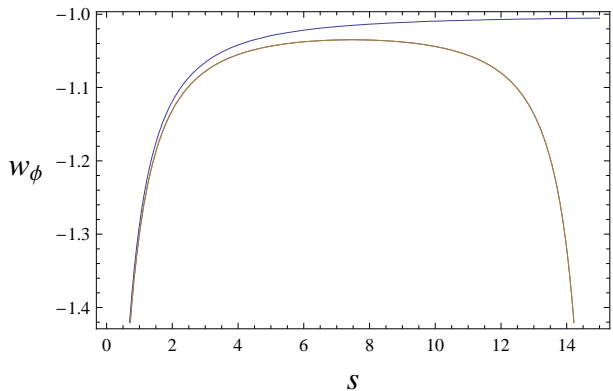,width=8cm,angle=0}} \caption{{\it
 $w$-evolution for the potential to $V=
V_0/\cosh(\phi/\phi_0)$, in terms of the variable $s=t/\phi_0$,
for $M_p=V_0=1$ and $W(\phi)=T(\phi)=1$. The top curve corresponds
to $\phi_0=1$ and the lower one to $\phi_0 = 10$.}} \label{cosh}
\end{center}
\end{figure}

For completeness, in fig.~\ref{cosh} we depict another example of
this solution sub-class, namely corresponding to $V=
V_0/\cosh(\phi/\phi_0)$, with $M_p=V_0=1$ and $W(\phi)=T(\phi)=1$.
As we can see the solution only appears to converge for $\phi_0=1$, and diverges for larger values.

\begin{figure}[ht]
\begin{center}
\mbox{\epsfig{figure=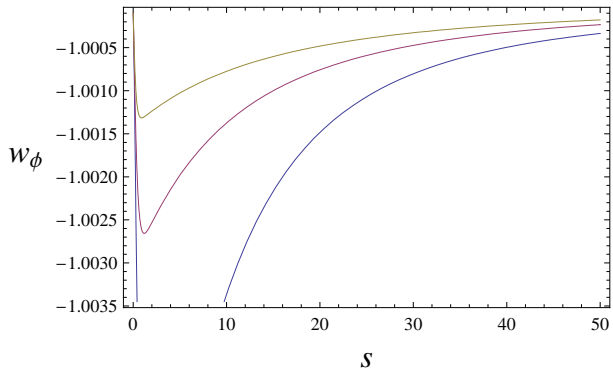,width=8cm,angle=0}} \caption{{\it
 $w$-evolution for $T(\phi)=
(\phi/L)^{\kappa}$, $V(\phi) = (\phi/\phi_0)^{\lambda}$ and
$W(\phi)=1$, in terms of the variable $s=t/\phi_0$, for $\epsilon
= \phi_0/L=0.1$ and $\kappa=\lambda=2$. The curves, from  bottom
to top, correspond to $\phi_0 = 10, 20, 30$, respectively.}}
\label{plot1}
\end{center}
\end{figure}

Let us now examine a more complicated case, considering $T(\phi)=
(\phi/L)^{\kappa}$ and $V(\phi) = (\phi/\phi_0)^{\lambda}$,
accompanied with $W(\phi)=1$. This choice introduces a new
mass-scale, which combined with $\phi_0$ allows us to write
$\epsilon = \phi_0/L$. If we desire to provide a theoretical
justification through supergravity solutions then $L$ is typically
large since it governs the radius of an $Ads$ spacetime, and thus $\epsilon$ will be small.
In fig.~\ref{plot1} we present the corresponding $w(s)$. One notices
that as $\phi_0$ is increased, the equation of state tends to $-1$
from below, however it is never too far away from $-1$. Finally,
numerical investigation reveals that the solution is sensitive to
the $\epsilon$-value, with smaller $\epsilon$ leading $w_{min}$ to
larger time-scales.

\section{Conclusions}
\label{conclusions}

In this work we have introduced a novel mechanism for realizing
either quintessence or phantom dark-energy-dominated phases,
within a string-theoretical context. This mechanism is based upon
the existence and subsequent dynamics of Ghost $Gp$-branes in a
warped compactification of the type IIB theory, which cancel the
combined BPS sectors of the $Dp$-brane, preserving the same
supersymmetries as their $Dp$ counterparts.

The scenario at hand admits a wide range of cosmological behavior,
depending on the various terms arising from the supergravity
background. In the simplest case, consisting of a single
$G3$-brane and with $V(\phi)/T(\phi)$ being constant, we see that
phantom behavior will dominate the phase-space dynamics for
sufficiently large $V(\phi)/T(\phi), \gamma$, since the
phase-space singularity $\gamma_c$ is pushed to larger values.
Beyond the singularity one finds a quintessence solution,
asymptotically tending towards $w_{\phi} = 0$. Although these
features arise from this particular model-subclass, it is clear
that more complicated behavior can be revealed considering more
general $W(\phi)$, $T(\phi)$, $V(\phi)$ cases, with a natural
realization of quintessence and phantom behavior, of the
$-1$-crossing and of a Big Rip.

Surprisingly enough, although the corresponding $Dp$-brane
scenario experiences only quintessence-type solutions
\cite{Gumjudpai:2009uy}, the present $Gp$-model may lead to both
quintessence or phantom cosmology. One can proceed to a more
detailed investigation of the phase-space behavior of $Gp$-brane
cosmological scenarios, for various cases of the involved tension
and potentials \cite{usfuture}. Alternatively one can impose the
desired cosmological evolution, and re-construct the corresponding
aforementioned functions. Since in this work we desire to remain
general, exploring the qualitative kinematic features of the
$Gp$-brane model, we do not proceed to such extensions, leaving
them for a future investigation \cite{usfuture}.

One remaining issue pertains to the quantum stability of such a
construction. As it is typical for phantom models, the energy is
unbounded from below leading to potential problems upon
quantization. However, since the $Gp$-brane is treated
semi-classically, we may hope that quantizing the open-string
modes with the appropriate boundary conditions may regularize the
theory. In particular, since we are dealing with a phantom scalar
field, all of the relevant energy conditions are violated. This
feature suggests that the phantom may be an unstable mode. To
verify the stability one must resort to a quantum
field-theoretical analysis. However, developing a quantum theory
of world-volume open-string modes using the DBI-action has proven
to be difficult, since the $D$-brane itself is a non-perturbative
state with regards to the string coupling. A boundary state
analysis may be possible, but it is beyond the scope of the
current work. Finally, we mention that since the usual phantom
models are robust only for small momenta (because at larger
momenta the higher-derivative terms dominate), one could estimate
the quasi-stable lifetime of the phantom field, provided the
momentum cut-off is fine-tuned and the phantom decays solely into
gravitons. Similarly, we could follow this line of reasoning  for
the model at hand, although this means that the field should decay
to the closed-string vacuum in a way that the open-string modes
give rise to gravitons. For the static case, where the $G3$-branes
are localised, they play a similar role to Orientifold planes. In
conclusion, we stress that the quantum stability of such
negatively-charged objects within string theory is still an open
question, but one that is ripe for future exploration.

We end this work referring to an additional advantage of the model
at hand, namely that it possesses a concrete UV-completion.
Therefore, it would be interesting to try to embed such branes
into full stringy compactifications, particularly if they could
serve as replacements for Orientifold planes. Obviously, the
underlying theory would then be $\mathcal{N}=2$, which is
phenomenologically unfavored, however there might be another
mechanism in the bulk which breaks half of this residual
supersymmetry. Exploring the nature of such scenarios is something
we leave for future endeavor.

\section*{Acknowledgments}
J.W is partly supported by NSERC of Canada.  E.N.S wishes to thank
Institut de Physique Th\'eorique, CEA, for the hospitality during
the preparation of the present work.


\begin{thebibliography}{0}


\bibitem{observations}
A.~G.~Riess {\it et al.}  [Supernova Search Team Collaboration],
Astron.\ J.\  {\bf 116}, 1009 (1998); S. Perlmutter {\it et al.}
[Supernova Cosmology Project Collaboration], Astrophys. J. {\bf
517}, 565 (1999).

\bibitem{quint}
B.~Ratra and P.~J.~E.~Peebles, Phys.\ Rev.\ D {\bf 37}, 3406
(1988); C.~Wetterich, Nucl.\ Phys.\ B {\bf 302}, 668 (1988);
A.~R.~Liddle and R.~J.~Scherrer, Phys.\ Rev.\ D {\bf 59}, 023509
(1999); I.~Zlatev, L.~M.~Wang and P.~J.~Steinhardt, Phys.\ Rev.\
Lett.\ {\bf 82}, 896 (1999).

\bibitem{phantBigRip}
R.~R.~Caldwell, M.~Kamionkowski and N.~N.~Weinberg, Phys. Rev.
Lett. {\bf 91}, 071301 (2003).


\bibitem{phant} R. R. Caldwell, Phys.
Lett. B {\bf{545}}, 23 (2002); S. Nojiri and S. D. Odintsov, Phys.
Lett. B {\bf 562}, 147 (2003); V. K. Onemli and R. P. Woodard,
Phys.\ Rev.\ D {\bf 70}, 107301 (2004);
  X.~m.~Chen, Y.~g.~Gong and E.~N.~Saridakis,
  JCAP {\bf 0904}, 001 (2009);
  E.~N.~Saridakis,
  Nucl.\ Phys.\  B {\bf 819}, 116 (2009).


\bibitem{quintom}
B.~Feng, X.~L.~Wang and X.~M.~Zhang, Phys.\ Lett.\  B {\bf 607},
35 (2005);
Z. K. Guo, {\it{et al.}}, Phys. Lett. B {\bf 608}, 177 (2005);
M.-Z Li, B. Feng, X.-M Zhang, JCAP, 0512, 002 (2005); B. Feng, M.
Li, Y.-S. Piao and X. Zhang, Phys. Lett. B {\bf 634}, 101 (2006);
W. Zhao and Y. Zhang, Phys. Rev. D {\bf73}, 123509 (2006);
  M.~R.~Setare and E.~N.~Saridakis,
  Phys.\ Lett.\  B {\bf 668}, 177 (2008);
  M.~R.~Setare and E.~N.~Saridakis,
  JCAP {\bf 0809}, 026 (2008).


\bibitem{ordishov}
 P. Bin\'{e}truy, C. Deffayet, D. Langlois, Nucl. Phys. B {\bf565}, 269 (2000);
G.R. Dvali, G. Gabadadze, M. Porrati, Phys. Lett. B {\bf485}, 208
(2000); S. Capozziello, Int. J. Mod. Phys. D {\bf11}, 483 (2002);
 S.Nojiri
and S.~D.~Odintsov, Phys. Rev. D {\bf{68}}, 123512 (2003);
  A.~Lue and G.~Starkman,
  Phys.\ Rev.\  D {\bf 67}, 064002 (2003);
  P.~S.~Apostolopoulos, N.~Brouzakis, E.~N.~Saridakis and N.~Tetradis,
  Phys.\ Rev.\  D {\bf 72}, 044013 (2005);
  G.~Calcagni, S.~Tsujikawa and M.~Sami,
  Class.\ Quant.\ Grav.\  {\bf 22}, 3977 (2005).

\bibitem{Horava}
  P.~Horava,
  Phys.\ Rev.\  D {\bf 79}, 084008 (2009);
  G.~Calcagni,
  arXiv:0904.0829 [hep-th];
  E.~Kiritsis and G.~Kofinas,
  Nucl.\ Phys.\  B {\bf 821}, 467 (2009);
  H.~Lu, J.~Mei and C.~N.~Pope,
  arXiv:0904.1595 [hep-th];
   C.~Charmousis, G.~Niz, A.~Padilla and P.~M.~Saffin,
  arXiv:0905.2579 [hep-th];
  E.~O.~Colgain and H.~Yavartanoo,
  arXiv:0904.4357 [hep-th];
  E.~N.~Saridakis,
  arXiv:0905.3532 [hep-th];
  X.~Gao, Y.~Wang, R.~Brandenberger and A.~Riotto,
  arXiv:0905.3821 [hep-th];
  M.~i.~Park,
  arXiv:0905.4480 [hep-th];
  M.~Botta-Cantcheff, N.~Grandi and M.~Sturla,
  arXiv:0906.0582 [hep-th];
  M.~i.~Park,
  arXiv:0906.4275 [hep-th];
 C. Bogdanos and E. N. Saridakis, arXiv:0907.1636 [hep-th].

\bibitem{Grana:2005jc}
  M.~Grana,
  Phys.\ Rept.\  {\bf 423}, 91 (2006).


\bibitem{Alishahiha:2004eh}
  M.~Alishahiha, E.~Silverstein and D.~Tong,
  Phys.\ Rev.\  D {\bf 70}, 123505 (2004);
  E.~Silverstein and D.~Tong,
  Phys.\ Rev.\  D {\bf 70}, 103505 (2004).

\bibitem{Kecskemeti:2006cg}
  S.~Kecskemeti, J.~Maiden, G.~Shiu and B.~Underwood,
  JHEP {\bf 0609}, 076 (2006);
  X.~Chen,
  Phys.\ Rev.\  D {\bf 71}, 063506 (2005);
  B.~Gumjudpai, T.~Naskar and J.~Ward,
  JCAP {\bf 0611}, 006 (2006);
  J.~Ward,
  JHEP {\bf 0712}, 045 (2007);
  X.~Chen,
  JCAP {\bf 0812}, 009 (2008);
  F.~Gmeiner and C.~D.~White,
  JCAP {\bf 0802}, 012 (2008).


\bibitem{Lidsey:2007gq}
  J.~E.~Lidsey and I.~Huston,
  JCAP {\bf 0707}, 002 (2007);
  D.~Baumann and L.~McAllister,
  Phys.\ Rev.\  D {\bf 75}, 123508 (2007).

\bibitem{Komatsu:2008hk}
  E.~Komatsu {\it et al.}  [WMAP Collaboration],
  Astrophys.\ J.\ Suppl.\  {\bf 180}, 330 (2009).

\bibitem{Easson:2007dh}
  D.~A.~Easson, R.~Gregory, D.~F.~Mota, G.~Tasinato and I.~Zavala,
  JCAP {\bf 0802}, 010 (2008).

\bibitem{Cai:2008if}
  Y.~F.~Cai and W.~Xue,
  arXiv:0809.4134 [hep-th].

\bibitem{Cai:2009hw}
  Y.~F.~Cai and H.~Y.~Xia,
  arXiv:0904.0062 [hep-th].

\bibitem{Thomas:2007sj}
  S.~Thomas and J.~Ward,
  Phys.\ Rev.\  D {\bf 76}, 023509 (2007);
  S.~Thomas and J.~Ward,
  JHEP {\bf 0610}, 039 (2006);
  S.~Thomas and J.~Ward,
  JHEP {\bf 0611}, 019 (2006).


\bibitem{Becker:2007ui}
  M.~Becker, L.~Leblond and S.~E.~Shandera,
  Phys.\ Rev.\  D {\bf 76}, 123516 (2007);
  T.~Kobayashi, S.~Mukohyama and S.~Kinoshita,
  JCAP {\bf 0801}, 028 (2008);
  S.~Mukohyama,
  Gen.\ Rel.\ Grav.\  {\bf 41}, 1151 (2009).




\bibitem{Silverstein:2008sg}
  E.~Silverstein and A.~Westphal,
  Phys.\ Rev.\  D {\bf 78}, 106003 (2008);
L.~McAllister, E.~Silverstein and A.~Westphal,
arXiv:0808.0706 [hep-th].


\bibitem{singlebrane}
  J.~Martin and M.~Yamaguchi,
  Phys.\ Rev.\  D {\bf 77}, 123508 (2008);
  Z.~K.~Guo and N.~Ohta,
  JCAP {\bf 0804}, 035 (2008).

\bibitem{Gumjudpai:2009uy}
  B.~Gumjudpai and J.~Ward,
  arXiv:0904.0472 [astro-ph.CO].


\bibitem{Myers:1999ps}
  R.~C.~Myers,
  JHEP {\bf 9912}, 022 (1999).

\bibitem{Myers:2003bw}
  R.~C.~Myers,
  Class.\ Quant.\ Grav.\  {\bf 20}, S347 (2003).

\bibitem{Brecher:2004qi}
  D.~Brecher, K.~Furuuchi, H.~Ling and M.~Van Raamsdonk,
  JHEP {\bf 0406}, 020 (2004).


\bibitem{Constable:1999ac}
  N.~R.~Constable, R.~C.~Myers and O.~Tafjord,
  Phys.\ Rev.\  D {\bf 61}, 106009 (2000).

\bibitem{Hyakutake:2004rh}
  Y.~Hyakutake,
  Phys.\ Rev.\  D {\bf 71}, 046007 (2005).

\bibitem{Tseytlin:1997csa}
  A.~A.~Tseytlin,
  Nucl.\ Phys.\  B {\bf 501}, 41 (1997);
  P.~Bordalo, L.~Cornalba and R.~Schiappa,
  Nucl.\ Phys.\  B {\bf 710}, 189 (2005).

\bibitem{Constable:2001ag}
  N.~R.~Constable, R.~C.~Myers and O.~Tafjord,
  JHEP {\bf 0106}, 023 (2001);
  P.~L.~H.~Cook, R.~de Mello Koch and J.~Murugan,
  Phys.\ Rev.\  D {\bf 68}, 126007 (2003).

\bibitem{Papageorgakis:2006ed}
  C.~Papageorgakis and S.~Ramgoolam,
  Int.\ J.\ Mod.\ Phys.\  A {\bf 21}, 6055 (2006).

\bibitem{Thomas:2006ac}
  S.~Thomas and J.~Ward,
  JHEP {\bf 0611}, 019 (2006).

\bibitem{McAllister:2007bg}
  L.~McAllister and E.~Silverstein,
  Gen.\ Rel.\ Grav.\  {\bf 40}, 565 (2008).

\bibitem{Langlois:2009ej}
  D.~Langlois, S.~Renaux-Petel and D.~A.~Steer,
  arXiv:0902.2941 [hep-th].

\bibitem{Dabholkar:1997zd}
  A.~Dabholkar,
  arXiv:hep-th/9804208.

\bibitem{Okuda:2006fb}
  T.~Okuda and T.~Takayanagi,
  JHEP {\bf 0603}, 062 (2006).

\bibitem{Terashima:2006qm}
  S.~Terashima,
  JHEP {\bf 0605}, 067 (2006).

\bibitem{Evans:2006eq}
  N.~Evans, T.~R.~Morris and O.~J.~Rosten,
  Phys.\ Lett.\  B {\bf 635}, 148 (2006).

\bibitem{Hao:2003ib}
  J.~g.~Hao and X.~z.~Li,
  Phys.\ Rev.\  D {\bf 68}, 043501 (2003);
  X.~z.~Li and J.~g.~Hao,
  Phys.\ Rev.\  D {\bf 69}, 107303 (2004).

\bibitem{Sen:1999md}
   A.~Sen,
   JHEP {\bf 9910}, 008 (1999);
  J.~Kluson,
  Phys.\ Rev.\  D {\bf 62}, 126003 (2000);
  E.~A.~Bergshoeff, M.~de Roo, T.~C.~de Wit, E.~Eyras and S.~Panda,
  JHEP {\bf 0005}, 009 (2000);
  M.~R.~Garousi,
  Nucl.\ Phys.\  B {\bf 584}, 284 (2000).

\bibitem{Vikman:2004dc}
  A.~Vikman,
  Phys.\ Rev.\  D {\bf 71}, 023515 (2005).

\bibitem{typeII*}
H. P. Nilles, Phys. Rept. 110, {\bf 1} (1984);
C. M. Hull, JHEP 0111, {\bf 012} (2001).



\bibitem{BigRip}
P.~F.~Gonzalez-Diaz, Phys.\ Rev.\ D 68, 021303 (2003); R.~Kallosh,
J.~Kratochvil, A.~Linde, E.~Linder and M.~Shmakova, JCAP
{\bf{0310}}, 015 (2003); G.~W.~Gibbons, [arXiv:hep-th/0302199].


\bibitem{Bounce}
  R.~Brandenberger,
  arXiv:0904.2835 [hep-th];
  R.~H.~Brandenberger,
  arXiv:0905.1514 [hep-th];
Y.~F.~Cai and E.~N.~Saridakis,
  arXiv:0906.1789 [hep-th].

\bibitem{expon}
P.G. Ferreira, M. Joyce, Phys. Rev. Lett.  {\bf79}, 4740 (1997);
E.~J.~Copeland, A.~R.~Liddle and D.~Wands,
 Phys.\ Rev.\  D {\bf 57}, 4686 (1998).

\bibitem{usfuture}
E.~N.~Saridakis and J.~Ward, in preparation.


\end{thebibliography}
\end{document}